\documentclass[prd,preprint,showpacs,preprintnumbers]{revtex4}

\usepackage{graphicx}
\usepackage{dcolumn}
\usepackage{bm}

\begin{document}


\title{A Search for Time-Dependent $B_s^0 - \overline{B_s^0}$ Oscillations \\
Using Exclusively Reconstructed $D^{\pm}_s$ Mesons}

%
\def\iAOMORI{$^{(1)}$}
\def\iBRI{$^{(2)}$}
\def\iBRUN{$^{(3)}$}
\def\iBU{$^{(4)}$}
\def\iCOLO{$^{(5)}$}
\def\iCSU{$^{(6)}$}
\def\iFERR{$^{(7)}$}
\def\iFRAS{$^{(8)}$}
\def\iJHU{$^{(9)}$}
\def\iLBL{$^{(10)}$}
\def\iMASS{$^{(11)}$}
\def\iMISSI{$^{(12)}$}
\def\iMIT{$^{(13)}$}
\def\iMOSCOW{$^{(14)}$}
\def\iNAGO{$^{(15)}$}
\def\iOREG{$^{(16)}$}
\def\iOXF{$^{(17)}$}
\def\iPERU{$^{(18)}$}
\def\iRAL{$^{(19)}$}
\def\iRUTG{$^{(20)}$}
\def\iSLAC{$^{(21)}$}
\def\iSOONG{$^{(22)}$}
\def\iTENN{$^{(23)}$}
\def\iTOHO{$^{(24)}$}
\def\iUCSB{$^{(25)}$}
\def\iUCSC{$^{(26)}$}
\def\iVAND{$^{(27)}$}
\def\iWASH{$^{(28)}$}
\def\iWISC{$^{(29)}$}
\def\iYALE{$^{(30)}$}

\author{
  \baselineskip=.75\baselineskip  
\mbox{Kenji Abe\unskip,\iNAGO}
\mbox{Koya Abe\unskip,\iTOHO}
\mbox{T. Abe\unskip,\iSLAC}
\mbox{I. Adam\unskip,\iSLAC}
\mbox{H. Akimoto\unskip,\iSLAC}
\mbox{D. Aston\unskip,\iSLAC}
\mbox{K.G. Baird\unskip,\iMASS}
\mbox{C. Baltay\unskip,\iYALE}
\mbox{H.R. Band\unskip,\iWISC}
\mbox{T.L. Barklow\unskip,\iSLAC}
\mbox{J.M. Bauer\unskip,\iMISSI}
\mbox{G. Bellodi\unskip,\iOXF}
\mbox{R. Berger\unskip,\iSLAC}
\mbox{G. Blaylock\unskip,\iMASS}
\mbox{J.R. Bogart\unskip,\iSLAC}
\mbox{G.R. Bower\unskip,\iSLAC}
\mbox{J.E. Brau\unskip,\iOREG}
\mbox{M. Breidenbach\unskip,\iSLAC}
\mbox{W.M. Bugg\unskip,\iTENN}
\mbox{T.H. Burnett\unskip,\iWASH}
\mbox{P.N. Burrows\unskip,\iOXF}
\mbox{A. Calcaterra\unskip,\iFRAS}
\mbox{R. Cassell\unskip,\iSLAC}
\mbox{A. Chou\unskip,\iSLAC}
\mbox{H.O. Cohn\unskip,\iTENN}
\mbox{J.A. Coller\unskip,\iBU}
\mbox{M.R. Convery\unskip,\iSLAC}
\mbox{R.F. Cowan\unskip,\iMIT}
\mbox{G. Crawford\unskip,\iSLAC}
\mbox{C.J.S. Damerell\unskip,\iRAL}
\mbox{M. Daoudi\unskip,\iSLAC}
\mbox{N. de Groot\unskip,\iBRI}
\mbox{R. de Sangro\unskip,\iFRAS}
\mbox{D.N. Dong\unskip,\iSLAC}
\mbox{M. Doser\unskip,\iSLAC}
\mbox{R. Dubois\unskip,\iSLAC}
\mbox{I. Erofeeva\unskip,\iMOSCOW}
\mbox{V. Eschenburg\unskip,\iMISSI}
\mbox{S. Fahey\unskip,\iCOLO}
\mbox{D. Falciai\unskip,\iFRAS}
\mbox{J.P. Fernandez\unskip,\iUCSC}
\mbox{K. Flood\unskip,\iMASS}
\mbox{R. Frey\unskip,\iOREG}
\mbox{E.L. Hart\unskip,\iTENN}
\mbox{K. Hasuko\unskip,\iTOHO}
\mbox{S.S. Hertzbach\unskip,\iMASS}
\mbox{M.E. Huffer\unskip,\iSLAC}
\mbox{M. Iwasaki\unskip,\iOREG}
\mbox{D.J. Jackson\unskip,\iRAL}
\mbox{P. Jacques\unskip,\iRUTG}
\mbox{J.A. Jaros\unskip,\iSLAC}
\mbox{Z.Y. Jiang\unskip,\iSLAC}
\mbox{A.S. Johnson\unskip,\iSLAC}
\mbox{J.R. Johnson\unskip,\iWISC}
\mbox{R. Kajikawa\unskip,\iNAGO}
\mbox{M. Kalelkar\unskip,\iRUTG}
\mbox{H.J. Kang\unskip,\iRUTG}
\mbox{R.R. Kofler\unskip,\iMASS}
\mbox{R.S. Kroeger\unskip,\iMISSI}
\mbox{M. Langston\unskip,\iOREG}
\mbox{D.W.G. Leith\unskip,\iSLAC}
\mbox{V. Lia\unskip,\iMIT}
\mbox{C. Lin\unskip,\iMASS}
\mbox{G. Mancinelli\unskip,\iRUTG}
\mbox{S. Manly\unskip,\iYALE}
\mbox{G. Mantovani\unskip,\iPERU}
\mbox{T.W. Markiewicz\unskip,\iSLAC}
\mbox{T. Maruyama\unskip,\iSLAC}
\mbox{A.K. McKemey\unskip,\iBRUN}
\mbox{R. Messner\unskip,\iSLAC}
\mbox{K.C. Moffeit\unskip,\iSLAC}
\mbox{T.B. Moore\unskip,\iMASS}
\mbox{M. Morii\unskip,\iSLAC}
\mbox{D. Muller\unskip,\iSLAC}
\mbox{V. Murzin\unskip,\iMOSCOW}
\mbox{S. Narita\unskip,\iTOHO}
\mbox{U. Nauenberg\unskip,\iCOLO}
\mbox{H. Neal\unskip,\iYALE}
\mbox{G. Nesom\unskip,\iOXF}
\mbox{N. Oishi\unskip,\iNAGO}
\mbox{D. Onoprienko\unskip,\iTENN}
\mbox{R.S. Panvini\unskip,\iVAND}
\mbox{C.H. Park\unskip,\iSOONG}
\mbox{I. Peruzzi\unskip,\iFRAS}
\mbox{M. Piccolo\unskip,\iFRAS}
\mbox{L. Piemontese\unskip,\iFERR}
\mbox{R.J. Plano\unskip,\iRUTG}
\mbox{R. Prepost\unskip,\iWISC}
\mbox{C.Y. Prescott\unskip,\iSLAC}
\mbox{B.N. Ratcliff\unskip,\iSLAC}
\mbox{J. Reidy\unskip,\iMISSI}
\mbox{P.L. Reinertsen\unskip,\iUCSC}
\mbox{L.S. Rochester\unskip,\iSLAC}
\mbox{P.C. Rowson\unskip,\iSLAC}
\mbox{J.J. Russell\unskip,\iSLAC}
\mbox{O.H. Saxton\unskip,\iSLAC}
\mbox{T. Schalk\unskip,\iUCSC}
\mbox{B.A. Schumm\unskip,\iUCSC}
\mbox{J. Schwiening\unskip,\iSLAC}
\mbox{V.V. Serbo\unskip,\iSLAC}
\mbox{G. Shapiro\unskip,\iLBL}
\mbox{N.B. Sinev\unskip,\iOREG}
\mbox{J.A. Snyder\unskip,\iYALE}
\mbox{H. Staengle\unskip,\iMASS}
\mbox{A. Stahl\unskip,\iSLAC}
\mbox{P. Stamer\unskip,\iRUTG}
\mbox{H. Steiner\unskip,\iLBL}
\mbox{D. Su\unskip,\iSLAC}
\mbox{F. Suekane\unskip,\iTOHO}
\mbox{A. Sugiyama\unskip,\iNAGO}
\mbox{S. Suzuki\unskip,\iNAGO}
\mbox{M. Swartz\unskip,\iJHU}
\mbox{F.E. Taylor\unskip,\iMIT}
\mbox{J. Thom\unskip,\iSLAC}
\mbox{E. Torrence\unskip,\iMIT}
\mbox{T. Usher\unskip,\iSLAC}
\mbox{J. Va'vra\unskip,\iSLAC}
\mbox{R. Verdier\unskip,\iMIT}
\mbox{D.L. Wagner\unskip,\iCOLO}
\mbox{A.P. Waite\unskip,\iSLAC}
\mbox{S. Walston\unskip,\iOREG}
\mbox{A.W. Weidemann\unskip,\iTENN}
\mbox{E.R. Weiss\unskip,\iWASH}
\mbox{J.S. Whitaker\unskip,\iBU}
\mbox{S.H. Williams\unskip,\iSLAC}
\mbox{S. Willocq\unskip,\iMASS}
\mbox{R.J. Wilson\unskip,\iCSU}
\mbox{W.J. Wisniewski\unskip,\iSLAC}
\mbox{J.L. Wittlin\unskip,\iMASS}
\mbox{M. Woods\unskip,\iSLAC}
\mbox{T.R. Wright\unskip,\iWISC}
\mbox{R.K. Yamamoto\unskip,\iMIT}
\mbox{J. Yashima\unskip,\iTOHO}
\mbox{S.J. Yellin\unskip,\iUCSB}
\mbox{C.C. Young\unskip,\iSLAC}
\mbox{H. Yuta\unskip.\iAOMORI}
\it
  \vskip \baselineskip                   
  \centerline{(The SLD Collaboration)}   
  \vskip \baselineskip        
  \baselineskip=.75\baselineskip   
\iAOMORI
  Aomori University, Aomori , 030 Japan, \break
\iBRI
  University of Bristol, Bristol, United Kingdom, \break
\iBRUN
  Brunel University, Uxbridge, Middlesex, UB8 3PH United Kingdom, \break
\iBU
  Boston University, Boston, Massachusetts 02215, \break
\iCOLO
  University of Colorado, Boulder, Colorado 80309, \break
\iCSU
  Colorado State University, Ft. Collins, Colorado 80523, \break
\iFERR
  INFN Sezione di Ferrara and Universita di Ferrara, I-44100 Ferrara, Italy, \break
\iFRAS
  INFN Lab. Nazionali di Frascati, I-00044 Frascati, Italy, \break
\iJHU
  Johns Hopkins University,  Baltimore, Maryland 21218-2686, \break
\iLBL
  Lawrence Berkeley Laboratory, University of California, Berkeley, California 94720, \break
\iMASS
  University of Massachusetts, Amherst, Massachusetts 01003, \break
\iMISSI
  University of Mississippi, University, Mississippi 38677, \break
\iMIT
  Massachusetts Institute of Technology, Cambridge, Massachusetts 02139, \break
\iMOSCOW
  Institute of Nuclear Physics, Moscow State University, 119899, Moscow Russia, \break
\iNAGO
  Nagoya University, Chikusa-ku, Nagoya, 464 Japan, \break
\iOREG
  University of Oregon, Eugene, Oregon 97403, \break
\iOXF
  Oxford University, Oxford, OX1 3RH, United Kingdom, \break
\iPERU
  INFN Sezione di Perugia and Universita di Perugia, I-06100 Perugia, Italy, \break
\iRAL
  Rutherford Appleton Laboratory, Chilton, Didcot, Oxon OX11 0QX United Kingdom, \break
\iRUTG
  Rutgers University, Piscataway, New Jersey 08855, \break
\iSLAC
  Stanford Linear Accelerator Center, Stanford University, Stanford, California 94309, \break
\iSOONG
  Soongsil University, Seoul, Korea 156-743, \break
\iTENN
  University of Tennessee, Knoxville, Tennessee 37996, \break
\iTOHO
  Tohoku University, Sendai 980, Japan, \break
\iUCSB
  University of California at Santa Barbara, Santa Barbara, California 93106, \break
\iUCSC
  University of California at Santa Cruz, Santa Cruz, California 95064, \break
\iVAND
  Vanderbilt University, Nashville,Tennessee 37235, \break
\iWASH
  University of Washington, Seattle, Washington 98105, \break
\iWISC
  University of Wisconsin, Madison,Wisconsin 53706, \break
\iYALE
  Yale University, New Haven, Connecticut 06511. \break
\rm
} 
%


\date{\today}

\begin{abstract}
A search for $B_s^0 - \overline{B_s^0}$ oscillations is performed using
a sample of 400,000 hadronic $Z^0$ decays collected
by the SLD experiment.
The $B_s^0$ candidates are reconstructed in the $B_s^0\rightarrow D_s^- X$ 
channel with $D_s^- \rightarrow \phi \pi^-$, $K^{*0}K^-$.
The $B_s^0$ production flavor is determined using the  
large forward-backward asymmetry of polarized 
$Z^0 \rightarrow b \overline{b}$ decays and charge information
in the hemisphere opposite that of the $B_s^0$ candidate.
The decay flavor is tagged by the charge of the $D_s^\pm$.
From a sample of 361 candidates with an average $B_s^0$ purity of
40\%, we exclude the following values of the oscillation frequency:
$\Delta m_s < 1.4$ ps$^{-1}$ and 
$2.4 < \Delta m_s < 5.3$ ps$^{-1}$ at the 95\% confidence level.
\end{abstract}



\maketitle


\section{Introduction}

The primary motivation for studying neutral $B$ meson oscillations 
is to measure the poorly known Cabibbo-Kobayashi-Maskawa (CKM) matrix element $V_{td}$.  
The $B_d^0$ oscillation frequency corresponds to 
the mass difference, $\Delta m_d$, between the physical eigenstates of the 
$B_d^0-\overline{B_d^0}$ system, 
which is sensitive to $|V_{td}|$.
Although $\Delta m_d$ is measured to within  
2.5\%~\cite{PDG2001}, theoretical uncertainties 
lead to a 20\% uncertainty in the extraction of 
$|V_{td}|$~\cite{Bernard:2000ki}. 
However, many uncertainties cancel in the ratio of mass 
differences in the 
$B_d^0$ and $B_s^0$ systems: 
\begin{equation}
{\Delta m_s\over\Delta m_d}={m_{B_s} f_{B_s}^2B_{B_s}\over m_{B_d} f_{B_d}^2B_{B_d}}
\left|{V_{ts}\over V_{td}}\right|^2=
(1.15\pm0.05)^2\left|{V_{ts}\over V_{td}}\right|^2,
\label{eqdelMdMs}
\end{equation}
where $m_{B_d}$ and $m_{B_s}$ are the $B$ meson masses, 
$f_{B_d}$ and $f_{B_s}$ are the decay constants, and 
$B_{B_d}$ and $B_{B_s}$ are the ``B-parameters''.
Using this formula, and assuming 
$|V_{ts}|$=$|V_{cb}|$,   
one can obtain
a 5-10\% theoretical uncertainty on $|V_{td}|$~\cite{Bernard:2000ki, Kronfeld:2002ab}.

As yet, the $B_s^0$ oscillation frequency has not been measured.
The published lower limit on $\Delta m_s$ based on the combined results from ALEPH, DELPHI,
OPAL and CDF is 13.1~ps$^{-1}$ at the 95\% confidence level~\cite{PDG2001}.
In the context of the Standard Model, other 
measurements suggest that $\Delta m_s$ may be just beyond this current 
limit~\cite{CIUCHINI}.

This letter describes a study~\cite{cjlthesis} of $B_s^0 - \overline{B_s^0}$ oscillations with the SLD experiment at SLAC.
The measurement of $B_s^0$ mixing requires 
excellent decay time resolution in order to resolve the
very fast oscillations.
The technique described in this letter,
using $B_s^0\rightarrow D_s^- X$ decays~\cite{conjugate}
with $D_s^-\rightarrow\phi \pi^-$ or $K^{*0}K^-$,
has excellent decay length 
resolution (the best of any $B_s^0$ analysis to date) and high purity 
(reconstruction of a $D_s^-$ helps to discriminate against the
prominent backgrounds from $B^+$ and $B_d^0$ mesons).
This work also represents the first use of polarized beams in
a search for $B_s^0$ mixing, which provide an effective new means of 
identifying the flavor of the $B_s^0$ meson at production. 

\section{Apparatus and Event Selection}

This analysis is based on a data set of 400,000 
events of the form $e^+e^-\rightarrow Z^0\rightarrow hadrons$,
collected from 1996 through 1998 by the SLD experiment. 
A detailed description of the experiment can be found 
elsewhere~\cite{SLDTDR,sldpaper}.
The analysis uses charged tracks reconstructed in the Central Drift Chamber 
(CDC) and the pixel-based CCD vertex detector (VXD3).  
The momentum resolution from the combined CDC and VXD3 fit is determined to be
$\sigma_{p\perp}/p_{\perp}=0.010\oplus 0.0024 p_{\perp}$, where 
$p_{\perp}$ (in GeV/c) is the momentum of the track transverse to the beamline.
The track impact parameter resolution in the transverse plane is 
$\sigma_{xy}=8\oplus 33/(p\sin^{3/2}\theta)$ $\mu m$ and along the
beam direction is 
$\sigma_{z}=10\oplus 33/(p\sin^{3/2}\theta)$~$\mu m$, where $p$ (in GeV/c) is 
the track momentum and $\theta$ is the polar angle with respect 
to the electron beam.  
The tracking system is surrounded by the Cherenkov Ring Imaging 
Detector (CRID), a two-radiator 
system that allows good pion and kaon separation in the momentum 
range between 0.3 and 35 GeV/c.

The goal of the analysis is to measure or constrain the oscillation frequency.
This is accomplished by using $B_s^0$ decay candidates that
are flavor tagged at both production and decay, and by measuring the proper time of
the decay through measurements of the $B_s^0$ decay length and energy.
The decay time distributions of $B$ mesons 
whose flavor at production and decay are different (the same),
so called mixed (unmixed) events, are modulated by the oscillation frequency. 
A fit to these distributions
constrains $\Delta m_s$.  

Event reconstruction consists of four main steps: hadronic
event selection, $Z^0\rightarrow b\overline b$ event selection, $D_s^-$ 
decay reconstruction,
and partial reconstruction of $B_s^0$ decays.
A hadronic event is identified as having at least seven charged tracks,
a total energy of at least 18 GeV, and an event thrust axis
satisfying $|cos\theta_{thrust}|<0.85$. The thrust axis is calculated based on
the energy clusters found in the liquid-argon calorimeter.  
The hadronic event selection removes essentially all
dilepton events and other non-hadronic backgrounds.
To enhance the fraction of $Z^0 \rightarrow b \overline{b}$
events in the sample, events are required to have at least 
one topologically reconstructed secondary vertex~\cite{bzvtop} with a vertex
mass greater than 2 GeV in either hemisphere. 
The vertex mass calculation includes a correction for the 
missing momentum transverse to the $B$ flight direction in order 
to partially account for missing particles.
A neural network~\cite{btom} 
is used to select a candidate secondary $B$ vertex (if multiple topological
secondary vertices exist) and its decay tracks.  
The resulting event sample is 97\% pure $b\overline{b}$ with a single 
hemisphere $b$ tagging efficiency of 54\%.

\begin{figure}
  \centering
  \includegraphics[width=3.4in]{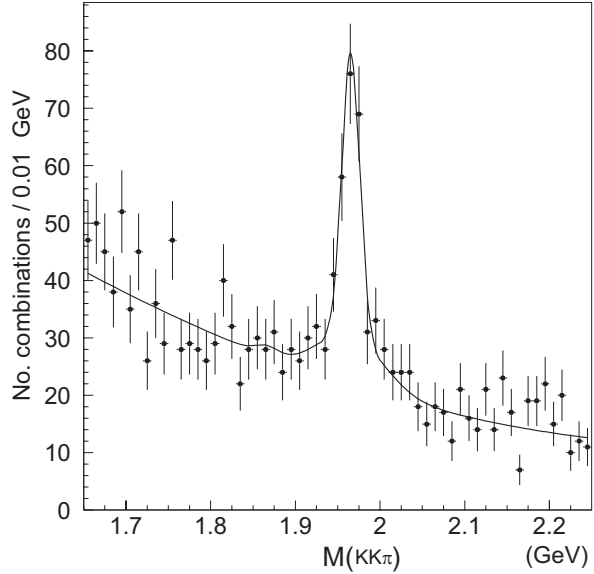}
  \caption{\label{fig:dsmass} Distribution of $KK\pi$ invariant mass for
  $\phi \pi$ and $K^{*0}K$ candidates combined, for the events 
  that include a $B_s^0$ candidate vertex. The solid line shows a fit
  with the same functional form as is used in the likelihood function:
  two Gaussian distributions with common mean for the $D_s$ signal, 
  a single Gaussian for the $D^+$
  contribution and a second-order polynomial for the background.
}  
\end{figure}

The $D_s^-$ is reconstructed in one of two modes,
$D_s^-\rightarrow\phi \pi^-~{\rm or}~K^{*0}K^-\rightarrow K^+K^-\pi^-$.
Oppositely charged tracks are first paired 
to form a $\phi$ ($K^{*0}$) candidate and a
third track is then attached to form a $D_s^-$ candidate.  
To maximize the discrimination between true $D_s^-$ and combinatorial 
background events, kinematic information for the $D_s^-$ candidate is fed 
into another neural network. In this case, the neural net inputs include 
CRID particle identification information for each of the three daughter tracks,
the $K^+K^-$ ($K^-\pi^+$) invariant mass for $\phi$ ($K^{*0}$) candidates,
the fit probability for the $D_s^-$ decay vertex,
the $D_s^-$ decay length normalized by the decay length error,
the total momentum of the $D_s^-$,
the angle between the neutral meson ($\phi$ or $K^{*0}$) momentum in the $D_s^-$ rest frame and the $D_s^-$ flight direction, and
the angle between
the $\pi^-$ or $K^-$ from the $D_s^-$ decay and the $K^+$ from the neutral meson decay in the rest frame of the neutral meson.
The neural net cut that maximizes the sensitivity of the
analysis to $B_s^0$ mixing is determined separately for each of the
two $D_s^-$ decay modes.
The resulting $KK\pi$ mass distribution for both modes combined is
shown in Figure~\ref{fig:dsmass}.


The $B_s^0$ decay vertex is found by vertexing the virtual $D_s^-$ track
with other tracks in the hemisphere.  
The virtual $D_s^-$ track is constructed by combining the 
4-momenta of the three daughter tracks and constraining
the parent $D_s^-$ to pass through the decay vertex. The parent track error matrix
is determined from propagation of the track measurement errors of the daughters.
The $B_s^0$ vertex fit is accomplished
in two steps: 1) identify an intersection of the virtual $D_s^-$ track with
another charged track in the same hemisphere to act as a seed
for the $B_s^0$ decay vertex and 2) add other charged tracks to
the seed vertex if they are more consistent with coming from the $B_s^0$
decay than coming from the primary interaction point (IP).
To find the 
seed, the $D_s^-$ track is individually vertexed with each track 
(excluding $D_s^-$ daughters)  
in the same hemisphere. 
The vertex that is farthest from the IP and upstream of the $D_s^-$ (or consistent
with being upstream within 5$\sigma$) and has a vertex 
fit $\chi^2$ of  
less than 5 is chosen as the seed.  
In order to determine if another track should be added to the vertex,
we examine two parameters: 1) the distance D from the IP to the seed vertex and
2) the distance L along the same direction from the IP to the point of closest approach of the track.
If the ratio L/D is greater than 0.5 and the track forms a good vertex with the
$D_s^-$ (fit $\chi^2 \le$ 5), the track is added to the vertex.
The latter condition
is imposed to reject spurious tracks (often from
a second charm decay if the $B$ decayed to two charm particles) that do
not point back to the $B$ vertex. 
Only $B$ decays for which the total charge of all associated tracks
is Q=0 or $\pm$1 are kept for the $\phi \pi$ mode and those for which
Q=0 are kept for the $K^{*0}K^-$ mode.
The selected tracks are then vertexed together 
with the $D_s^-$ 
to obtain the best estimate of the $B$ decay position.

The IP location, needed to calculate the $B_s^0$ decay length,
is determined from tracks that have vertex detector hits and 
extrapolate within 
3$\sigma$ of the beamline.
The coordinates transverse to the beam are averaged over approximately
30 sequential hadronic $Z^0$ events,
while the position along the beam is determined
event-by-event.  The resulting error in IP position is 
3.5$\mu m$ in the transverse plane and 17$\mu m$ along the beam, the best
resolution of any high energy physics collider.
By making use of this well-determined beam spot, the very small beam size, 
and information from the high precision CCD vertex detector, 
this analysis has a unique sensitivity
for measuring decay times of the $B$ mesons.

An estimate of the $B_s^0$ decay length resolution is determined 
event-by-event according to the vertex fit and IP uncertainties, with 
correction factors applied 
for each particle decay hypothesis ($B_s^0$, $B_d^0$, $B^+$ or $b$-baryon)
as determined from Monte Carlo simulation.  
The SLD Monte Carlo uses JETSET 7.4 with the $B$ decay 
model tuned to CLEO and ARGUS data~\cite{Chou:2001gj}.  
Parameterizing the decay length resolution as a sum of two Gaussians with
normalizations of 0.6 (core) and 0.4 (tail), 
the estimated multiplicative correction factors for the $B_s^0$ signal events 
are 1.07 and 2.16 for the core and tail resolutions, respectively. 
The resulting average decay length resolution for the $B_s^0$ signal events
is 50 $\mu m$ (core) and 151 $\mu m$ (tail).

The $B$ meson boost is 
calculated from separate estimates for the charged and neutral
particle contributions to the total energy.
The charged energy is determined by summing all the charged
tracks associated with the $B$ decay assuming the pion mass (except
for the two kaons from the $D_s^-$ decay).  The neutral 
energy estimate uses five different techniques.  The first
four techniques are calorimeter-based and use various
constraints (beam energy, jet energy, $B_s^0$ mass and calorimeter
information) to estimate
the neutral energy of the $B$ meson~\cite{bmoore}.
The fifth technique is based only on the information from the charged
decay tracks and the kinematics of the decay
($B$ vertex axis, charged track momentum and $B_s^0$ mass 
constraint)~\cite{bdong}. 
The results from the five algorithms are then averaged, taking correlations
into account, to obtain the total $B$ energy.
The resulting average boost resolution 
(${\sigma_{\gamma\beta}\over\gamma\beta}$) is represented by 
a sum of two Gaussians with widths (normalizations) of 8\% and 18\% 
(0.6 and 0.4) for the $B_s^0$ signal events.

Our final data sample includes 361 events within $\pm 40$MeV of the 
$D_s^-$ mass peak (Figure~\ref{fig:dsmass}) with an
average $D_s^-$ purity of 48.1\%.  
The composition of the 
$D_s^-$ signal sample is calculated from the published branching ratio
measurements with the relative reconstruction efficiencies of various
decay modes taken from the Monte Carlo simulation.  We estimate that
the $D_s^-$ signal peak consists of $B_s^0\rightarrow D_s^{\pm}X$ (55.1\%), 
$B_d^0\rightarrow D_s^{\pm}X$ (22.4\%), 
$B^+\rightarrow D_s^{\pm}X$ (15.6\%), 
\hbox{$b$\thinspace -baryon$\rightarrow D_s^{\pm}X$} (5.5\%) and 
prompt $\overline{c} \rightarrow D_s^-$
(1.4\%).  For the hadronic $B_s^0$ decays, 
roughly 10\% of the decays yield a wrong-sign $D_s$ 
($B_s^0\rightarrow D_s^+ X$ instead of 
$B_s^0\rightarrow D_s^- X$), due to 
$W^+\rightarrow D_s^+$.  Of the 361 $B_s^0$ candidates, 39 are semileptonic
decay candidates ($B_s^0\rightarrow D_s^- l^+ \nu X$).  The 
fraction of wrong-sign $D_s$ decays
for the semileptonic modes is about 5\%.  These events have
significantly higher $B_s$ fraction and tagging purity than the hadronic
decay sample, and therefore are parameterized separately.

\section{Tagging}

The flavor of the $B_s^0$ at decay is determined by the charge of the
$D_s^-$. A $D_s^-$ is assumed to come from a $B_s^0$ and a 
$D_s^+$ is assumed to come from a $\overline{B_s^0}$.
The $B_s^0$ flavor at production is obtained by exploiting 
the large forward-backward asymmetry in polarized $e^+e^-_{L,R} \rightarrow
Z^0\rightarrow b\bar{b}$ decays.  The differential cross section for 
the decay is given by 
\begin{equation}
\frac{d\sigma(b)}{d{\cos\theta}}\propto (1-A_e P_e)(1+\cos^2\theta)+
2A_b(A_e-P_e)\cos\theta,
\label{eqdiffcross}
\end{equation}
where the asymmetry parameter can be expressed in terms of vector and 
axial-vector couplings, $A_f=2a_{f}v_{f}/(a_{f}^2+v_{f}^2)$ and 
$\theta$ is defined as the angle between the outgoing fermion and 
the electron direction.
The electron polarization is defined as 
$P_e=\frac{N(R)-N(L)}{N(R)+N(L)}$,
where $N(R)$ ($N(L)$) is the number of right-handed (left-handed) electrons
in a beam bunch.
The outgoing $b$-quark is produced preferentially along the 
direction opposite to the spin of the $Z^0$ boson.  Therefore,
by knowing the polarization of the electron beam and the direction of the
jet, the flavor of the primary quark in the jet can be inferred.
The correct tag probability depends on the polar angle of
the jet with respect to the incident electron beam direction and the 
electron beam polarization.
The average electron beam polarization achieved during the run is about 73\%.
The resulting average correct tag probability is about 72\%.
In addition to the polarization tag, information from the
hemisphere opposite to the reconstructed $B_s^0$ is used to improve
the identification of the production flavor.  
A series of neural networks is used to
combine momentum-weighted jet charge, vertex charge, 
dipole charge~\cite{sldpaper}, lepton charge and kaon 
charge information.
The purity of the opposite hemisphere charged tags has been calibrated
from the data to be (70.4$\pm$0.8)\%.
Combining all available tags, the overall production flavor tag purity is 
(77.8$\pm$0.8)\%.  In the end, candidates for which the flavor of the $B$ meson at
decay is more than 50\% likely to be different from the flavor at 
production are said to be `mixed'.

\section{Fitting and Results}

To determine the $B_s^0$ oscillation frequency, an unbinned likelihood function
is used to describe the proper time distribution of mixed and unmixed events.
For signal $B_s^0\rightarrow D_s^\pm X$ 
events tagged as mixed (unmixed), the proper time distribution has the form, 
\begin{eqnarray}
{P}_{B_s}(\Delta m_s,t_{rec})^{mixed}_{unmixed}=
\int_{-\infty}^{+\infty}\frac{1}{2\tau_{B_s}}
\exp(-t/\tau_{B_s})&[&1\mp (1-2\eta_{B_s})\cos(\Delta m_s t)] \cdot \nonumber \\ 
&&\epsilon_{B_s}(t)\cdot G_{B_s}(t_{rec},t)\cdot dt, 
\label{eqBslike}
\end{eqnarray}
where $t_{rec}$ is the reconstructed proper time, $t$ is the true proper
time, $\tau_{B_s}$ is the $B_s^0$ lifetime, $\epsilon_{B_s}(t)$ is the 
reconstruction efficiency and 
$G_{B_s}(t_{rec},t)$ is the proper time resolution function for the 
$B_s^0$ events.  The overall mistag probability is
$\eta_{B_s}=\eta_i(1-\eta_{B_s}^f) + (1-\eta_i)\eta_{B_s}^f$, where 
$\eta_i$ is the production flavor mistag probability and $\eta_{B_s}^f$ is
the decay flavor mistag probability for the $B_s^0$ events.
The efficiency and proper time resolution
functions are derived from Monte Carlo simulations.
The Monte Carlo vertexing resolution is based on 
our understanding of the impact parameter 
resolution, which is carefully tuned to match the data.  
A comparison of decay length resolution
between data and Monte Carlo using 3-prong $\tau$ decays shows 
good agreement.
The proper time resolution is expressed 
in terms of 
decay length and boost resolutions ($\sigma_L$,$\sigma_{\gamma\beta}$) 
according to:
\begin{equation}
\sigma_{t}^{ij}(t) = \left[\left(\frac{\sigma_L^i}{\gamma\beta c}\right)^2
+ \left(t\,\frac{\sigma_{\gamma\beta}^j}{\gamma\beta}\right)^2\right]^{1/2}
\label{eqn:sigmat}\ ,
\end{equation}
where the indices $i$, $j$ =1 for core and 2 for tail resolution.  Both
$\sigma_L$ and $\sigma_{\gamma\beta}$ are determined event-by-event.
The decay length resolution
$\sigma_L$  depends on the $B_s^0$ vertex fit error matrix and 
IP uncertainties, and the boost resolution $\sigma_{\gamma\beta}$ 
is parameterized as a function
of charged track energy in the decay.
The decay time resolution function is comprised of four $\sigma_{t}$ 
Gaussian distributions given by
the various $\sigma_L$ and $\sigma_{\gamma\beta}$ core-tail combinations.

The distribution for $B_d^0$ mesons, which are also subject to
oscillations, is identical to equation~\ref{eqBslike} 
with $B_s^0$ subscripts replaced by $B_d^0$ subscripts.
The terms for $B^+$ and $b$-baryon, which do not oscillate, are constructed 
in an analogous fashion, 
but with $\Delta m=0$.
The probability density function 
for the data sample has the form,
\begin{eqnarray}
{P}&=&{\it f_{D_s}}\left(
\frac{\it f_{B_s}}{N_{B_s}}{P}_{B_s}+
\frac{\it f_{B_d}}{N_{B_d}}{P}_{B_d}+
\frac{\it f_{B^+}}{N_{B^+}}{P}_{B^+}+
\frac{\it f_{b-bary}}{N_{b-bary}}{P}_{b-bary}+
\frac{\it f_{cc}}{N_{cc}}{{P}}_{cc}\right) \nonumber \\
&+&\left[1-{\it f_{D_s}}\right]{P}_{comb}.
\label{eqfulllike}
\end{eqnarray}
The fraction of $D_s^-$ signal above the combinatorial
$m_{KK\pi}$ background, ${\it f_{D_s}}$, is estimated from the
previous $D_s^-$ mass fit (Figure~\ref{fig:dsmass}) 
as a function of $m_{KK\pi}$.  
The remaining fractions are 
calculated based on the measured branching ratios with reconstruction 
efficiencies taken from the 
Monte Carlo simulation.  The function ${P}_{cc}$
describes the prompt charm ($Z^0\rightarrow c\bar{c}$)
events, and is obtained from Monte Carlo simulation.
The time distribution for the combinatorial events, ${P}_{comb}$ is
parameterized directly from the data using events in the $D_s^-$ 
mass sidebands.
The sideband regions are defined as: 1.7 $<$ $m_{KK\pi}$ $<$ 1.8~GeV 
(lower sideband) and 2.05 $<$ $m_{KK\pi}$ $<$ 2.2~GeV (upper sideband).  
The parameters $N_i$, are normalization constants, obtained by 
integrating the sum of $P_{mixed}$ and $P_{unmixed}$ over all reconstructed
proper time.  

In the absence of a signal, the amplitude fit method~\cite{Moser} 
is used to set a limit on $\Delta m_s$.  The amplitude fit is 
equivalent to a Fourier analysis, in which one searches for peaks
in the frequency spectrum of oscillation. To perform an amplitude fit, the likelihood function is modified by
replacing $\cos(\Delta m_s t)$ with $A\cdot \cos(\Delta m_s t)$.  The 
amplitude $A$ and its error $\sigma_A$ are 
then measured at each assumed value of 
$\Delta m_s$.  
If mixing occurs at the chosen value 
of $\Delta m_s$, the fitted value of $A$ should be consistent with unity. 
At values of $\Delta m_s$ sufficiently far from the true mixing
frequency, the fitted value of A should 
be close to zero, consistent with no oscillation.
Values of $\Delta m_s$ for which $A+1.645\sigma_A\leq1$ 
can be excluded at the 95\% confidence level. We have tested this
procedure on simulated data for $\Delta m_s$ values of 
4, 10, 17 and 270 ps$^{-1}$ to verify that the amplitude fit behaves as
expected in all cases.

\begin{figure}
   \centering
   \includegraphics[width=3.4in]{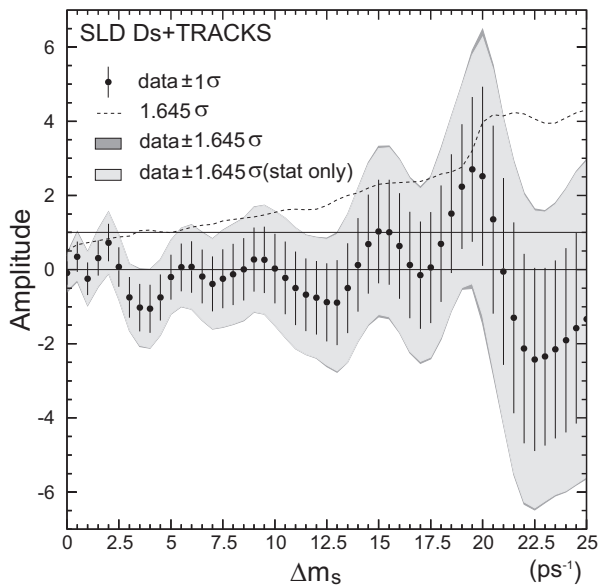}
   \caption{\label{fig:ampfit}
   Measured $B_s^0$ oscillation amplitude as a function of $\Delta m_s$.
    The light-grey (dark-grey) band shows the 90\%
   confidence region obtained from statistical (total) uncertainties.
   Values of $\Delta m_s$ for which the allowed band is below $A=1.0$ 
   are excluded at the one-sided 95\% confidence level.}
\end{figure}

The amplitude plot for this analysis is shown in Figure~\ref{fig:ampfit}.
There is no evidence for a significant signal anywhere in the plot, so 
we use the analysis to set a limit on $\Delta m_s$.
The systematic errors on $A$ are evaluated according to
reference~\cite{Moser}, all of which are negligible compared to
the statistical errors.
The systematics are dominated by uncertainties in a small
reconstructed proper time bias (evaluated at 100\% of the 
simulated correction, typically a few hundreths of a picosecond), 
in the production flavor tag (evaluated with $\pm0.8$\% uncertainty
in the average tag probability and a reweighting of the shape of the
distribution as a function of neural net output), 
in $f_{D_s^-}$ (as determined
from the $KK\pi$ mass fit), and in the boost resolution (10\%).
To a lesser extent, there are contributions from uncertainties in decay length
resolution (7\%) and branching fractions assumed in the fit. Uncertainties
in particle lifetimes and in the $B_d^0$ oscillation frequency
are completely negligible. The list of systematic errors is shown in 
Table~\ref{tbl:syst_err}.
The dark-grey band in Figure~\ref{fig:ampfit}, barely visible at the
edge of the light-grey band, shows the effect of adding the total
systematic error to the calculations.

\begin{table*}
\caption{\label{tbl:world_br} $B$ production fractions and various 
branching ratios assumed in the amplitude fit. The uncertainties for the
branching ratios do not include the uncertainty from 
${\cal B}$($D_s\rightarrow~\phi\pi$).}
\begin{center}
\begin{ruledtabular}
\begin{tabular}{ccc} 
Parameter &Value and Error&Ref. \\  \hline
{\it f}($\overline{b} \rightarrow B_s^0$) &
 $0.100\pm 0.012 $ &\cite{WORKG} \\
{\it f}($\overline{b} \rightarrow B_d^0, B^+$) & 
$0.401\pm 0.010$&\cite{WORKG} \\
{\it f}($\overline{b} \rightarrow b-baryon$) & 
$0.099\pm 0.017$&\cite{WORKG} \\ 
\hline
$R_b \cdot {\it f}(b\rightarrow \overline{B_s^0}) \cdot 
{\cal B}(\overline{B_s^0}
 \rightarrow D_s^+X)\cdot {\cal B}(D_s^+\rightarrow \phi \pi^+)$&
$(6.21^{+0.71}_{-0.78})\times 10^{-4}$ & \cite{WORKG,ALEPH1} \\
${\it f}(b\rightarrow W^- \rightarrow D_s^-)\cdot 
{\cal B}(D_s^- \rightarrow \phi \pi^-)$&
$(3.66\pm 0.45)\times 10^{-3}$ & \cite{ALEPH1} \\
${\cal B}(B_d^0,B^+\rightarrow D_s^{\pm} X)\cdot 
{\cal B}(D_s^- \rightarrow \phi \pi^-)$ &
$(3.71\pm 0.28)\times 10^{-3}$ & \cite{WORKG} \\
${\cal B}(B_d^0,B^+\rightarrow D_s^{-} X) / 
{\cal B}(B_d^0,B^+\rightarrow D_s^{\pm} X)$ &
$0.172\pm 0.083 $ & \cite{WORKG} \\ 
${\cal B}(\overline{c} \rightarrow D_s^-)\cdot 
{\cal B}(D_s^- \rightarrow \phi \pi^-)$&
$(3.4\pm 0.3)\times 10^{-3}$ & \cite{WORKG} \\ 
\end{tabular}
\end{ruledtabular}
\end{center}
\end{table*}

\begin{table*}
\caption{\label{tbl:syst_err}
Table of statistical and dominant systematic uncertainties 
for several $\Delta m_s$ values.}
\begin{ruledtabular}
\begin{tabular}{cccc}
 $\Delta m_s$            & ~~10 ps$^{-1}$  & ~~15 ps$^{-1}$
                         & ~~20 ps$^{-1}$ \\
 \hline
 Measured amplitude $A$  &       ~0.029  &     $1.027$
                         &      $ 2.513$ \\
$\sigma_A^{stat}$  &
    $\pm 0.933$  &   $\pm 1.361$
                         &   $\pm 2.283$  \\
$\sigma_A^{syst}$       &
         $^{+0.088}_{-0.084}$ & $^{+0.312}_{-0.313}$
       & $^{+0.776}_{-0.789}$ \\
 \hline
  {\it f}($\overline{b} \rightarrow B_s^0$) &
         $^{+0.012}_{-0.012}$ &
         $^{+0.021}_{-0.022}$ & $^{+0.010}_{-0.011}$ \\
  {\it f}($\overline{b} \rightarrow b-$baryon) &
         $^{+0.004}_{-0.004}$   &
         $^{+0.001}_{-0.001}$ & $^{-0.019}_{+0.019}$ \\
  $R_b \cdot {\cal B}(b\rightarrow \overline{B_s^0})\cdot
   {\cal B}(\overline{B_s^0}\rightarrow D_s^+X) $ &
         $^{-0.032}_{+0.043}$ &
         $^{-0.040}_{+0.048}$ & $^{-0.068}_{+0.068}$ \\
 ${\cal B}(b\rightarrow W^- \rightarrow D_s^-)$  &
         $^{+0.013}_{-0.013}$ &
         $^{+0.022}_{-0.022}$ & $^{+0.010}_{-0.012}$ \\
 ${\cal B}(B_d^0,B^+\rightarrow D_s^{\pm} X)$   &
         $^{+0.013}_{-0.014}$ &
         $^{+0.013}_{-0.014}$ & $^{+0.042}_{-0.047}$ \\
  $\frac{{\cal B}(B_d^0,B^+\rightarrow D_s^{-} X)}{
{\cal B}(B_d^0,B^+\rightarrow D_s^{\pm} X)}$  &
         $^{+0.012}_{-0.013}$ &
         $^{+0.058}_{-0.060}$ & $^{+0.153}_{-0.156}$ \\
 ${\cal B}(\overline{c} \rightarrow D_s^-)$  &
         $^{+0.003}_{-0.003}$ &
         $^{+0.007}_{-0.007}$ & $^{+0.011}_{-0.011}$ \\
 Decay length resolution      &
         $^{+0.034}_{-0.036}$ &
         $^{+0.015}_{-0.015}$ & $^{+0.026}_{-0.045}$ \\
 Boost resolution      &
         $^{+0.049}_{-0.048}$ &
         $^{-0.052}_{+0.042}$ & $^{-0.161}_{+0.140}$ \\
 ${\it f}_{D_s}$      &
         $^{+0.019}_{-0.018}$ &
         $^{+0.096}_{-0.096}$ & $^{-0.090}_{+0.076}$ \\
 Average production flavor tag  &
         $^{-0.029}_{+0.031}$ &
         $^{-0.036}_{+0.036}$ & $^{-0.051}_{+0.048}$ \\
 Production flavor tag shape  &
         $^{-0.020}_{+0.018}$ &
         $^{+0.008}_{-0.013}$ & $^{+0.306}_{-0.318}$ \\
 Proper time offset     &
         $^{+0.007}_{-0.007}$ &
         $^{+0.278}_{-0.278}$ & $^{+0.671}_{-0.671}$ \\
\end{tabular}
\end{ruledtabular}
\end{table*}

Based on the result of the amplitude fit, the
values of the $B_s^0$ oscillation frequency excluded at the 95\% confidence level
are: 
$\Delta m_s < 1.4$ ps$^{-1}$ and 
$2.4 < \Delta m_s < 5.3$ ps$^{-1}$.

\section{Conclusions}
In conclusion, we have performed a search for $B_s^0 - \overline{B_s^0}$ oscillations 
using 361 $B_s^0\rightarrow D_s^- X$ candidate events with an average
$B_s^0$ purity of 40\%.  
The excluded values of oscillation frequency are:
$\Delta m_s < 1.4$ ps$^{-1}$ and 
$2.4 < \Delta m_s < 5.3$ ps$^{-1}$ at the 95\% confidence level.
The analysis exploits the unique large polarized
forward-backward asymmetry in $Z^0 \rightarrow b\bar{b}$ decays to
enhance the production flavor tag.
Combining the good production flavor tag and excellent proper time resolution, 
the analysis contributes to the world average at high $\Delta m_s$ despite low statistics.
Taken in conjunction with other published results, our result 
raises the 95\% C.L. world limit on $\Delta m_s$ from 13.1~ps$^{-1}$
to 13.9~ps$^{-1}$.

\section*{Acknowledgments}
        We thank the personnel of the SLAC accelerator department and
the technical staffs of our collaborating institutions for their outstanding
efforts.  This work was supported by the Department of Energy, the National
Science Foundation, the Instituto Nazionale di Fisica of Italy, the
Japan-US Cooperative Research Project on High Energy Physics, and the
Science and Engineering Research Council of the United Kingdom.


\begin{thebibliography}{99}
\bibitem{conjugate}
Charge 
conjugate reactions are implied throughout this paper.
\end{thebibliography}
\end{document}